%% file: dfp_com_aware.tex
\documentclass[10 pt, conference]{ieeeconf}

\IEEEoverridecommandlockouts                              
\usepackage{bbold}
\usepackage{color}
\usepackage{graphics} 
\input{my_sections.tex}

\input{mysymbol.sty}
\usepackage{theorem}
\usepackage{cite}
\usepackage{amsmath}
\usepackage{amssymb}
\usepackage{mathtools}
\usepackage{multirow}
\usepackage{url}
\usepackage{graphics, subfigure, times, amsfonts}
\usepackage{tikz, epic,eepic}
\usetikzlibrary{shapes,arrows}
\usepackage{pgfplots}
\usepackage{color}
\usepackage{hyperref}
\usepackage{epstopdf}
\usepackage{epsfig, amsbsy}
\usepackage{latexsym}
\usepackage{amscd, verbatim}
\usepackage{multirow}
\usepackage{booktabs}

\usepackage{algorithm}
\usepackage{algorithmic}

\renewcommand{\baselinestretch}{0.97}\selectfont
\newtheorem{theorem}{Theorem}

\newtheorem{lemma}{Lemma}
\newtheorem{definition}{Definition}

{\itshape}{\rmfamily}
\newtheorem{assumption}{Assumption}

\def\forall{\text{for all\ }}

\title{\LARGE 
Decentralized Fictitious Play Converges Near a Nash Equilibrium in Near-Potential Games}
\author{Sarper Ayd\i n, Sina Arefizadeh and Ceyhun Eksin %
\thanks{S. Aydin and C. Eksin are with the Industrial and Systems Engineering Department, Texas A\&M University, College Station, TX 77843. E-mail:{\tt\small  \; sarper.aydin@tamu.edu; eksinc@tamu.edu.} This work was supported by NSF CCF-2008855.
}
\thanks{Sina Arefizadeh was with the Industrial and Systems Engineering Department, Texas A\&M University, College Station, TX 77843.}}

\begin{document}
\normalsize
\maketitle

%
\begin{abstract}
We investigate convergence of decentralized fictitious play (DFP) in near-potential games, wherein agents preferences can almost be captured by a potential function. In DFP agents keep local estimates of other agents' empirical frequencies, best-respond against these estimates, and receive information over a time-varying communication network. We prove that empirical frequencies of actions generated by DFP converge around a single Nash Equilibrium (NE) assuming that there are only finitely many Nash equilibria, and the difference in utility functions resulting from unilateral deviations is close enough to the difference in the potential function values. This result assures that DFP has the same convergence properties of standard Fictitious play (FP) in near-potential games. 
\end{abstract}

%

%
\section{Introduction}

\input{introduction.tex}
\section{Near-Potential Games}\label{sec:model}
%
In a game $\Gamma$, agents defined by the set $\ccalN=\{1,\cdots,N\}$, select actions over a common finite set $a_i \in \ccalA_i=\ccalA$, where $i \in \ccalN$. Agents have individual utility values $u_i: \ccalA^N \rightarrow \reals$ corresponding to joint action profiles $(a_i,a_{-i}) \in \ccalA^N$ where $-i:=\{ j \in \ccalN \setminus \{i\}\}$ denotes the set of all agents except $i$. The game is formally defined by the tuple $\Gamma:=(\ccalN,\ccalA^N, \{u_i\}_{i \in \ccalN}\}$. Potential games can be defined as follows \cite{monderer1996potential}.

\begin{definition}[Potential Games] \label{def_potential}
A game $\Gamma$ is a potential game, if there exists a  potential function $u: \ccalA^N \rightarrow \mathbb{R}$ such that the following relation holds for all agents $i \in \ccalN$,
\begin{equation}\label{eq_def_potential}
    u(a'_i,a_{-i})-u(a_i,a_{-i})= u_i(a'_i,a_{-i})-u_i(a_i,a_{-i})
\end{equation}
where $a_i' \in \ccalA$ and $a_i\in \ccalA$ and $a_{-i} \in \ccalA_{-i}$.
\end{definition}


Potential function $u:\ccalA^N \rightarrow \reals$ mirrors the change in individual utilities $u_i:\ccalA^N \rightarrow \reals$ as a result of unilateral changes between two actions $a_i' \in \ccalA$ and $a_i \in \ccalA$ taken by each agent $i$.
We are going to define the class of near-potential games, using a metric of closeness between games in the following.

\begin{definition}[Maximum Pairwise Difference] \label{def_mpd}
Let $\Gamma=(\ccalN,\mathcal{A}^N,\{u_i\}_{i\in\ccalN})$ and  $\hat{\Gamma}=(\ccalN,\mathcal{A}^N, \hat{u}_i\}_{i\in\ccalN})$ be two  games with the same set of agents $\ccalN$ and the joint action sets $\mathcal{A}^N$ but possibly different utilities $\{u_i\}_{i \in \ccalN}$ and $\{\hat{u}_i\}_{i \in \ccalN}$. Further, let $d_{(a'_i,a)}^{\Gamma}:=u_i(a'_i,a_{-i})-u_i(a_i,a_{-i})$ be the unilateral change in utility of an agent $i$ by switching to an action $a'_i \in \ccalA_i$, given joint action profile $a=(a_i,a_{-i}) \in \ccalA^N$ in a game $\Gamma$. Then, the maximum pairwise difference $d(\Gamma, \hat{\Gamma})$ between the games $\Gamma$ and $\hat{\Gamma}$ is defined as, 
\begin{equation}\label{eq_mpd}
    d(\Gamma, \hat{\Gamma}):=  \underset{i \in \ccalN, \, a'_i \in \ccalA,\, a  \in \ccalA^N }{\max} |d_{(a'_i,a)}^{\Gamma}- d_{(a_i',a)}^{\hat{\Gamma}}|.
\end{equation}
\end{definition}
The maximum pairwise difference (MPD) defines the distance between two games based on unilateral changes of actions. Using MPD we define near-potential games as in \cite{candogan2013dynamics}.

\begin{definition} [Near-Potential Games]
A game $\Gamma$ is a near-potential game if there exists a potential game $\hat{\Gamma}$ within a maximum-pairwise distance (MPD), $d(\Gamma, \hat{\Gamma}) \le \delta$ where $\delta \in \mathbb{R}^+$.
\end{definition}

Near potential games are another relaxation of potential games. Other well-known extensions of potential games are ordinal \cite{monderer1996potential}, weighted \cite{monderer1996potential}, and best-response potential games \cite{voorneveld2000best}. Finding the potential function of near-potential games is studied at \cite{candogan2011flows} and a convex mathematical program is proposed. In this paper, our goal is to show the convergence of the decentralized FP in near-potential games.



%

\section{Decentralized Fictitious Play}\label{sec:DFP}
Fictitious play is an individual decision-making rule, where agents are assumed to select actions according to a stationary distribution (strategy) $\sigma_{i} \in \Delta \ccalA$, where $\Delta \ccalA$ is the set of probability distributions over the action set $\ccalA$. From now on, we suppose that each action $a \in \ccalA$ is represented with an unit vector $\bbe_k \in \{0,1\}^K$ where $|\ccalA|=K$. The stationary distribution of agent $i$ at time $t$, denoted with $f_{i,t} \in \Delta\ccalA$, is given by the empirical frequency of past actions taken by agent $i$, 
\begin{equation}\label{eq_empirical_frequency}
f_{i,t}= \frac{t-1}{t}f_{i,t-1}+\frac{1}{t}a_{i,t},
\end{equation}

The individual actions of agent $i$ is a result of it best responding to the stationary action distribution of other agents $f_{-i,t-1}:=\{f_{j,t-1}\}_{j\in\ccalN\setminus i}$,
\begin{equation}\label{eq_br}
    a_{i,t} \in \arg \max_{ a_i \in \ccalA_i} u_i(a_{i}, f_{-i,t-1}),
\end{equation}
where $u_i: \Delta\ccalA^N \rightarrow \reals$ can be defined as expected utility function on the set of probability distributions $\Delta \ccalA^N$ with little abuse of notation.  Given this definition, it also holds $(f_{i,t},f_{-i,t}) \in \Delta \ccalA^N$ and $a_i \in \Delta\ccalA$, as $a_i$ also becomes a degenerate distribution giving probability $1$ to a selected action. 

In a decentralized setting, it is not possible that the past actions of agent $j\in \ccalN \setminus i$ ($f_{j,t}$) is  available to agent $i$. An extension to FP assumes agents communicate over a network and exchange their beliefs about the empirical frequencies of other agents \cite{aydin2021decentralized}.

In particular, we assume agents communicate over a time-varying network $\ccalG=(\ccalN,\ccalE_t)$, where each agent $i$ only communicates with its current neighbors $\ccalN_{i,t} := \{j: (i,j)\in \ccalE_t\}$. In such a case, we replace the $f_{j,t}\in \Delta \ccalA$ with a local copy (belief) $\upsilon_{j,t}^i \in \Delta\ccalA_j$ kept at agent $i$. Given $\upsilon_{i,t}^i = f_{i,t}$ agent $i$ updates its local estimate with the local copies from its neighbors,
\begin{equation}\label{eq_local_update}
    \upsilon_{j,t}^i= \sum_{l \in \ccalN_i \cup \{i\} } w_{jl,t}^i \upsilon_{j,t}^{l}.
\end{equation}
where $w_{jl,t}^i \ge 0$ is the weight that agent $i$ puts on agent $l$'s estimate of agent $j$ such that $w_{j,l}^i \ge \eta$ for some $\eta>0$ only if $l\in \ccalN_i\cup \{i\}$, otherwise $w_{j,l}^i=0$, and $\sum_{l \in \ccalN_i\cup \{i\} }  w_{j,l}^i=1$ for all $i,j$. 

In DFP, agent $i$ selects an action $a_{i,t}$ to maximize its expected utility computed using its local beliefs $\upsilon_{-i,t-1}^i:=\{\upsilon_{j,t-1}^{i}\}_{j\in\ccalN\setminus i}$,
\begin{equation}\label{eq_br_local}
    a_{i,t} \in \arg \max_{ a_i \in \ccalA_i} u_i(a_{i}, \upsilon_{-i,t-1}^i).
\end{equation}
%
We summarize the steps of DFP algorithm below.

\begin{algorithm}[H] 
   \caption{DFP for Agent $i$}
\label{suboptimal_alg_inner}
\begin{algorithmic}[1]\label{alg_DFP}
   \STATE {\bfseries Input:} Local estimates $\upsilon_{-i0}^i$ and time-varying networks $\{\ccalG_t=(\ccalN,\ccalE_t)\}_{t\ge1}$.
\FOR{$t=1,2,\cdots $} 
    \STATE Select an action $a_{it}$ \eqref{eq_br_local} and  update $f_{i,t}$ \eqref{eq_empirical_frequency}. 
    \STATE Share and update local copies $\upsilon_{j,t}^i$ \eqref{eq_local_update}.
  \ENDFOR 
   \end{algorithmic}
\end{algorithm}

In the next section,  we are going to show that joint empirical frequencies $f_t=(f_i,f_{-i,t}) \in \Delta \ccalA^N$ converges around only one Nash Equilibrium (NE) of the closest potential game.

\section{Convergence of DFP in Near-Potential Games} \label{sec::conv}

\subsection{Preliminaries}
The joint strategy profile $\sigma^*$ is an approximate NE, if agent $i$ can obtain at most $\epsilon\ge0$ utility value, by changing its strategy to another strategy.
\begin{definition} [Approximate Nash Equilibrium] \label{def_app_Nash}
The joint strategy profile $\sigma^*=(\sigma_i^*,\sigma_{-i}^*) \in \Delta \ccalA^N$ is a $\epsilon$-Nash equilibrium of the game $\Gamma$ for $\epsilon\geq 0$ if and only if for all $i\in\ccalN$,
\begin{equation} \label{eq_app_Nash}
    u_i(\sigma^*_i,\sigma^*_{-i}) - u_i(\sigma_i,\sigma_{-i}^*) \ge -\epsilon, \quad \forall \sigma_i \in \Delta\ccalA_i.
\end{equation}
\end{definition}
We denote the set of $\epsilon$-Nash equilibria in a game $\Gamma$ with the notation $\Sigma_\epsilon$. If $\epsilon=0$ in \eqref{eq_app_Nash}, then $\sigma^*$ is a NE strategy profile. Next, we provide the notion of upper semi-continuity.

\begin{definition}[Upper Semi-continuous Correspondence]\label{def_upsc_cr}
A correspondence $h: X \Rightarrow Y$ is upper semi-continuous, if one of the following statements hold,
\begin{itemize}
    \item For any $\bar{x} \in X$ and any open neighborhood $V$ of $h(\bar{x})$, there exists a neighborhood $U$ of $\bar{x}$, such that $h(x) \subset V$, and $h(x)$ is a compact set for all $x \in U$.
    \item $Y$ is compact, and the set, \textit{i.e.} its graph, $\{(x,y)| x \in X, y \in h(x)\}$ is closed. 
\end{itemize}
\end{definition}

\subsection{Convergence Analysis}

We state the assumption on the structure of the time-varying communication network $\{\ccalG_t\}_{t\ge1}$ in the following.
\begin{assumption}\label{as_connect}
Time-varying communication networks $\{\ccalG_t\}_{t\ge1}$ satisfy the following assumptions,
\begin{enumerate}
    \item[\textit{i)}] The network $\ccalG=(\ccalN,\ccalE_{\infty})$ is connected, where $\ccalE_{\infty}=\{(i,j)| (i,j) \in \ccalE_t, \, \text{for infinitely many t} \in \naturals \}$.

    \item[\textit{ii)}] There exists a time step $T_{B}>0$, such that for any edge $(i,j) \in \ccalE_{\infty}$ and $t \ge 1$, it holds  $(i,j) \in \bigcup_{\tau=0}^{T_{B}-1}\ccalE_{t+\tau}$.
\end{enumerate}
\end{assumption}
Assumption \ref{as_connect} $i)-ii)$ are referred as \textit{connectivity} and \textit{bounded communication interval} in order. This assumptions assures that any information about an agent $j$ reaches agent $i$ in some bounded time. 

\begin{assumption}\label{as_com_weights}
There exists a scalar $0<\eta<1$, such that the followings hold for all $i\in \ccalN$, $j\in \ccalN$ and $t=1,2,\dots$,
\begin{itemize}
    \item[\textit{(i)}] If $l \in \ccalN_{i,t} \cup  \{i\}$, then $w_{jl,t}^i \ge \eta$. Otherwise,  $w_{jl,t}^i=0$,
    \item[\textit{(ii)}] $w_{ii,t}^i =1$,  
    \item [\textit{(iii)}] $\sum_{l \in \ccalN_{i,t} \cup \{i\} } w_{jl,t}^i=1$.
\end{itemize}
\end{assumption}
Assumption \ref{as_com_weights}{\it (i)} indicates that agents place positive weight on estimates they receive from their current neighbors in \eqref{eq_local_update}. Assumption \ref{as_com_weights}{\it (ii)} guarantees  $\nu^i_{i,t}=f_{i,t}$ for all $t>0$. Assumption \ref{as_com_weights}{\it(iii)} assures that the weights $W_{j,t}$, defined as the collective update weights on agent $j$'s empirical frequency  $[W_{j,t}]_{i,l}=w_{jl,t}^i$, is row stochastic for all times. Now, we state the convergence of local estimates $\nu^i_{j,t}$  to empirical frequencies $f_{j,t}$---see  \cite{arefizadeh2019distributed} for the proof.
\begin{lemma}[Proposition 1,  \cite{arefizadeh2019distributed}]\label{lem_ups_to_fic}
Suppose Assumptions \ref{as_connect}-\ref{as_com_weights} hold. If $f_{j0}=\upsilon^i_{j,0}$ holds for all pairs of agents {$j\in \ccalN$ and $i\in \ccalN$}, then the local copies $\{\upsilon^i_{t}\}^{i \in \ccalN}_{ t \ge 0}$ converge to the empirical frequencies $\{f_{t}\}_{ t \ge 0}$ with rate $O(\log t /t )$, i.e., $|| \upsilon^i_{j,t} -f_{j,t} || = O(\log t /t )$ for all $j\in \ccalN$ and $i\in \ccalN$.
\end{lemma}

The proof mainly exploits the properties of row-stochastic matrices as per Assumption \ref{as_com_weights}. Next result provides the difference potential value of empirical frequencies between consecutive time steps---see \cite{aydin2021decentralized} for the proof.

\begin{lemma}[Lemma 2,  \cite{aydin2021decentralized}]\label{lem_inc}
Suppose Assumptions \ref{as_connect}-\ref{as_com_weights} hold. Let $\Gamma$ be a $\delta$ near-potential game for some $\delta\geq 0$. The potential function is given by $u(\cdot)$. We denote the empirical frequency sequence generated by the DFP algorithm as $\{f_t\}_{t \ge 1}$. If the empirical frequency $f_t$ is outside the $\epsilon$-NE set for $\epsilon \ge 0$, then given a long enough $T>0$ we have 
\begin{equation} \label{eq_step_inc}
    u({f_{t+1}})-u({f_{t}})\ge \frac{\epsilon-N\delta}{t+1} -O\Big(\frac{\log t}{t^2}\Big) \; \text{ for all } t\geq T.
\end{equation}
\end{lemma}
 Lemma \ref{lem_inc} suggests that after long enough time, if the empirical frequencies are outside the approximate NE region $N\delta$, the potential value of a close potential game increases. We next characterize the potential change if empirical frequencies follow an excursion path where they go outside of approximate-NE regions $N\delta+\epsilon_1$ and $N\delta+\epsilon_2$ in order, and return back firstly to $N\delta+\epsilon_2$ and then $N\delta+\epsilon_1$ given $0< \epsilon_1 <\epsilon_2 $.

\begin{lemma}\label{lem_step3}
 Suppose Assumptions \ref{as_connect}- \ref{as_com_weights} hold  Let $\{f_t\}_{t \ge 1}$ be the sequence generated by Algorithm \ref{alg_DFP}. Further, let $T_1, T_2, T'_2, T'_1$ be time steps such that $T<T_1 \le T_2 <T'_2 \le T'_1$, for large enough  $T>0$ defined as follows,
\begin{itemize}
 \item $T_1$ is a time step that holds $f_{T_1-1} \in \Sigma_{N\delta+\epsilon_1}$ and $f_{t} \not \in \Sigma_{N\delta+\epsilon_1}$, for all $T_1 \le t < T'_1$,
 \item $T_2$ is a time step that holds $f_{T_2-1} \in \Sigma_{N\delta+\epsilon_2}$ and $f_{t} \not \in \Sigma_{N\delta+\epsilon_2}$, for all $T_2 \le t < T'_2$,
 \item $T'_2$ is a time step that holds $f_{T'_2-1} \not \in \Sigma_{N\delta+\epsilon_2}$ and $f_{T_2'}  \in \Sigma_{N\delta+\epsilon_2}$,
 \item $T'_1$ is a time step that holds $f_{T'_1-1} \not \in \Sigma_{N\delta+\epsilon_1}$ and $f_{T_1'}  \in \Sigma_{N\delta+\epsilon_1}$,
\end{itemize}
where $\epsilon_1>0$, $\epsilon_2>0$. Then, there exist $0< \epsilon_1 <\epsilon_2 $ such that the following holds,
\begin{equation} \label{eq_lower_T1}
    u(f_{T_1'})-u(f_{T_1}) \ge \sum_{t=T_2}^{T_2'-1}\frac{2\epsilon_2}{3(t+1)}.
\end{equation}
\end{lemma}
\begin{proof}
From \eqref{eq_step_inc}, if $f_t \not \in \Sigma_{N\delta+\epsilon_2}$, i.e., $T< T_2 \le t < T'_2$, the following holds,
\begin{equation}\label{eq_eps2}
    u(f_{t+1})-u(f_t)\ge \frac{\epsilon_2}{(t+1)}-O \Big (\frac{\log t}{t^2}\Big) \ge \frac{2\epsilon_2}{3(t+1)}.
\end{equation}
Then, \eqref{eq_eps2} implies,
\begin{equation}\label{eq_eps2_sum}
    u(f_{T_2'})-u(f_{T_2})=\sum_{t=T_2}^{T_2'-1} u(f_{t+1})-u(f_t) \ge \sum_{t=T_2}^{T_2'-1}\frac{2\epsilon_2}{3(t+1)}.
\end{equation}
Similarly, for the time intervals, $T_1 \le t < T_2$ or
$T'_2 \le t < T'_1$, the following inequalities also hold,
\begin{equation}\label{eq_eps1_sum}
    u(f_{T_1'})-u(f_{T_2'})=\sum_{t=T'_2}^{T_1'-1} u(f_{t+1})-u(f_t) \ge \sum_{t=T_2'}^{T_1'-1}\frac{2\epsilon_1}{3(t+1)},
\end{equation}
\begin{equation}\label{eq_eps1_sum_2}
    u(f_{T_2})-u(f_{T_1})=\sum_{t=T_1}^{T_2-1} u(f_{t+1})-u(f_t) \ge \sum_{t=T_1}^{T_2-1}\frac{2\epsilon_1}{3(t+1)}.
\end{equation}
Thus, the result in \eqref{eq_lower_T1} follows by   \eqref{eq_eps2_sum}, \eqref{eq_eps1_sum}, \eqref{eq_eps1_sum_2}, and the fact that $u(f_{T'_1})-u(f_{T_1})=(u(f_{T'_1})-u(f_{T_2'}))+(u(f_{T'_2})-u(f_{T_2}))+(u(f_{T_2})-u(f_{T_1}))$.
\end{proof}

This result puts a lower bound on the excursion away from an approximate NE. We need the following additional assumptions to hold for the main convergence result.

\begin{assumption}\label{as_finite_eq}
The game $\Gamma:=(\ccalN,\mathcal{A}^N,\{u_i\}_{i\in\ccalN})$ has only a nonempty set of finitely many Nash equilibria, $\Sigma_0=\{\sigma^{*(1)},\sigma^{*(2)},\cdots,\sigma^{*(M)}\}$ for $M \in \mathbb{Z}^+$. 
\end{assumption}

Assumptions \ref{as_finite_eq} asserts that there needs to be only a finite number of Nash equilibria.

\begin{lemma}[Theorem 5.2, \cite{candogan2013dynamics} ]\label{lem_step1}
Suppose Assumption \ref{as_finite_eq} holds. Let $q: \mathbb{R}_+ \to \mathbb{R}_+$ be a function defined as follows,
\begin{equation}\label{eq_max_min}
    q(\alpha)= \underset{\sigma \in \Sigma_{\alpha}}{\max} \underset{m \in \{1,\cdots, M\}}{\min} || \sigma-\sigma^{*(m)}||,
\end{equation}
where $\sigma^{*(m)}$ is a NE of the game $\Gamma$ as defined in Assumption \ref{as_finite_eq}. Then, the function $q(\cdot)$ is \textit{i)} weakly increasing, \textit{ii)} upper semi-continuous, \textit{iii)}  satisfies $q(0)=0$ and $\underset{\alpha \rightarrow 0}{\lim}q(\alpha)= 0$.

\end{lemma}

The function $q: \mathbb{R}_+ \to \mathbb{R}_+$ defines the largest distance between the set of approximate Nash equilibria and the set of Nash equilibria. 
\begin{assumption}\label{as_dist_ne} 
Let $q: \mathbb{R}_+ \rightarrow \mathbb{R}_+$ be the function as defined in \eqref{eq_max_min}. The maximum pairwise distance between two games $d(\Gamma,\hat{\Gamma}) \le \delta < \bar{\delta}$ is small enough such that there exists $\bar{\alpha} >0$ that satisfies $ N\delta < N\bar{\delta} < \bar{\alpha}/2$ and $q(\bar{\alpha}) < d^*/4$, where $d^*$ is the minimum distance between any two equilibria, i.e., $d^*=\underset{m'\neq m''} {\min}||\sigma^{*(m')}-\sigma^{*(m'')}||$ , where $m', m'' \in \{1,\cdots, M\}$ .
\end{assumption}

Assumption \ref{as_dist_ne} provides a relation between the minimum distance between two different Nash equilibria of the near potential game, and the distance to a given potential game. Next, we state the main theorem of this study.

\begin{theorem}\label{lem_step4}
 Suppose Assumptions \ref{as_connect}- \ref{as_dist_ne} hold. Let $\{f_t\}_{t \ge 1}$ be the sequence generated by Algorithm \ref{alg_DFP}. The empirical frequencies $\{f_t\}_{t\ge1}$ converge to an approximate equilibrium set around a single equilibrium point, after long time enough $t>T$.
\end{theorem}
\begin{proof}
Suppose that $\epsilon_1=\bar{\epsilon}$ and $\epsilon_2=\bar{\alpha}-N\bar{\delta}$ such that $0<\epsilon_1<\epsilon_2$ is satisfied as stated in Lemma \ref{lem_step3}. Further, using $\bar{\delta}>\delta$ by Assumption \ref{as_dist_ne}, it holds $N\delta+\epsilon_1<N\delta+\epsilon_2<\bar{\alpha}=N\bar{\delta}+\epsilon_2$. Then, by this relation, it also holds, $\Sigma_{N\delta+\epsilon_1} \subset \Sigma_{N\delta+\epsilon_2} \subset  \Sigma_{N\bar{\delta}+\epsilon_2}$. Therefore, the sets $\Sigma_{N\delta+\epsilon_1}$ and $\Sigma_{N\delta+\epsilon_2}$ consist of disjoint neighborhoods of finitely many equilibrium points by Step 2 of Theorem 5.2 \cite{candogan2013dynamics}. This implies that for any $\sigma \in \Sigma_{N\delta+\epsilon_1}$  or $\sigma \in \Sigma_{N\delta+\epsilon_2}$, the definition of $q$ provides $|| \sigma-\sigma^{*(m)}|| \le q(N\delta+\epsilon_1)$ or $|| \sigma-\sigma^{*(m)}|| \le q(N\delta+\epsilon_2)$ in order, for exactly one equilibrium $\sigma^{*(m)}$. \\

To prove the statement, we are going to use a contradiction. Given the definitions of $T_1,T_2,T'_1,T'_2$ as in Lemma \ref{lem_step3}, we assume that the empirical frequencies $f_t$ leaves from $N\delta+\epsilon_1$-NE set around a equilibrium point $\sigma^{*(m')}$, and then enters again into $N\delta+\epsilon_1$-NE set, but around another equilibrium point $\sigma^{*(m'')}$ so that the following relations are given as below,
\begin{align}\label{eq_set}
    &f_{T_1-1} \in \Sigma_{N\delta+\epsilon_1} \;\; &\text{and} \;\; ||f_{T_1-1}-\sigma^{*(m')}|| \le q(N\delta+\epsilon_1),\\
    &f_{T_1'} \in \Sigma_{N\delta+\epsilon_1} \;\; &\text{and} \;\; ||f_{T_1'}-\sigma^{*(m'')}|| \le q(N\delta+\epsilon_1),\\
    &f_{T_2-1} \in \Sigma_{N\delta+\epsilon_2} \;\; &\text{and} \;\; ||f_{T_2-1}-\sigma^{*(m')}|| \le q(N\delta+\epsilon_2),\label{eq_T_2-1}\\
    &f_{T_2'} \in \Sigma_{N\delta+\epsilon_2} \;\; &\text{and} \;\; ||f_{T_2'}-\sigma^{*(m'')}|| \le q(N\delta+\epsilon_2)\label{eq_T_2'}.
\end{align}
Given \eqref{eq_T_2-1} and \eqref{eq_T_2'}, $d^*=\min_{m'\neq m''}||\sigma^{*(m')}-\sigma^{*(m'')}|| $, and the fact that $\bar{\alpha} > N\delta+\epsilon_2$ implies $q(N\delta+\epsilon_2) < q(\bar{\alpha}) < d^*/4 $, the distance between empirical frequencies $f_{T_2'}$ and $f_{T_2-1}$ can be bounded below 
\begin{equation}\label{eq_T2_dist}
    || f_{T_2'}-f_{T_2-1}|| > \frac{d^*}{2}.
\end{equation}
Next, we obtain an upper bound on the distance between empirical frequencies at consecutive time steps, using the update rule \eqref{eq_empirical_frequency} and triangle inequality,
\begin{equation}\label{eq_dif_f}
    ||f_{t+1}-f_t||= \frac{1}{t+1}||f_t-a_t|| \le (||f_t||+||a_t||) \le \frac{2N}{t+1}
\end{equation}
since it holds $||a_{i,t}|| \le1$ and $||f_{i,t}|| \le 1$ for all $i \in \mathcal{N}$ . Then, \eqref{eq_T2_dist} suggests $|| f_{T_2'}-f_{T_2}|| > \frac{d^*}{2}$ also holds, as $||f_{T_2}-f_{T_2-1}||$ is sufficiently small for a sufficiently large time instant $T_2>T$. Together, upper and lower bounds from $T_2$ to $T_2'$ can be provided as,
\begin{align} \label{eq_sum_T2}
    &\sum_{t=T_2}^{T_2'-1}\frac{2N}{t+1} \ge  \sum_{t=T_2}^{T_2'-1}||f_{t+1}-f_t||, \nonumber\\ 
    &\ge ||(\sum_{t=T_2}^{T_2'-1} f_{t+1}-f_t)|| = ||f_{T_2'}-f_{T_2}|| > \frac{d^*}{2}.
\end{align}
Since $\sum_{t=T_2}^{T_2'-1}\frac{2N}{t+1} > \frac{d^*}{2}$, it holds that $\sum_{t=T_2}^{T_2'-1} \frac{d^*(t+1)}{4N} <1$ dividing $\frac{d^*}{2}$ by  $\sum_{t=T_2}^{T_2'-1}\frac{2N}{t+1}$. Then, again multiplying $\sum_{t=T_2}^{T_2'-1}\frac{2\epsilon_2}{3(t+1)}$ by $\sum_{t=T_2}^{T_2'-1} \frac{d^*(t+1)}{4N}$, we obtain the following lower bound, 
\begin{equation}\label{eq_sum_T1}
     u(f_{T'_1})-u(f_{T_1}) \ge \sum_{t=T_2}^{T_2'-1}\frac{2\epsilon_2}{3(t+1)} \ge \frac{\epsilon_2 d^*}{6N}.
\end{equation}
For the next part, we are going to derive the lower bound on the difference between maximum and minimum potential values in the neighborhood of different NE points $\sigma^{*(m')}$ and $\sigma^{*(m'')}$. Let $\underline{\sigma}^{*(m')}=\arg \min_{\sigma \in || \sigma-a^{*(m')}|| \le N\delta+\epsilon_1} u(\sigma)$ and $\bar{\sigma}^{*(m'')}=\arg \max_{\sigma \in || \sigma-a^{*(m')}|| \le N\delta+\epsilon_1} u(\sigma)$ be the respective mixed action profiles that provides maximum and minimum values around two arbitrarily different equilibrium points. The difference between their potentials $u(\underline{\sigma}^{*(m')})$  and $u(\bar{\sigma}^{*(m'')})$ can be rewritten as,
\begin{align}\label{eq_dif_pot}
   &u(\underline{\sigma}^{*(m')})- u(\bar{\sigma}^{*(m'')}) \nonumber\\
   &=(u(\underline{\sigma}^{*(m')})- u(f_{T_1'}))+(u(f_{T_1'})-u(f_{T_1})) \nonumber \\
   &+(u(f_{T_1})-u(\bar{\sigma}^{*(m'')})).
\end{align}
For each segment of (RHS) of \eqref{eq_dif_pot}, we can derive a bound. Using Lipschitz continuity and \eqref{eq_set}-\eqref{eq_T_2'}, it gives $u(\underline{\sigma}^{*(m')})-u(f_{T_1'}) \ge -2Lq(N\delta+\epsilon_1)$ and $u(f_{T_1})-u(\bar{\sigma}^{*(m'')})\ge -2Lq(N\delta+\epsilon_1)$, where $L$ is Lipschitz constant of utility functions. Moreover, using again Lipschitz continuity and \eqref{eq_dif_f}, it follows that $u(f_{T_1})-u(f_{T_1-1})\ge -\frac{2NL}{T_1}$. Therefore, using \eqref{eq_sum_T1}, the following is provided as below,
\begin{equation}\label{eq_dif_equ}
    u(\underline{\sigma}^{*(m')})- u(\bar{\sigma}^{*(m'')}) \ge \frac{\epsilon_2 d}{6N}-4Lq(N\delta+\epsilon_1)-\frac{2NL}{T_1}.
\end{equation}
The values $\epsilon_2=\bar{\alpha}-N\bar{\delta}$, $\epsilon_1=\bar{\epsilon}$ can be replaced. Hence, given the facts that $\bar{\delta} > \delta$, and the function $q$ is increasing, \eqref{eq_dif_equ} takes the form,
\begin{equation}
    u(\underline{\sigma}^{*(m')})- u(\bar{\sigma}^{*(m'')}) \ge \frac{(\bar{\alpha}-N\bar{\delta}) d^*}{6N}-4Lq(N\bar{\delta}+\bar{\epsilon})-\frac{2NL}{T_1}.
\end{equation}
We also have that $ q( N\delta+\bar{\epsilon}) < \frac{(\bar{\alpha}- N\delta)d^*}{24NL}$ (see Theorem 5.2 in \cite{candogan2013dynamics}). This yields the fact that $u(\underline{\sigma}^{*(m')})- u(\bar{\sigma}^{*(m'')}) > 0$, after long enough time $T_1>T$, as $\frac{2NL}{T_1} \rightarrow 0$. Then, \eqref{eq_dif_f} and \eqref{eq_sum_T2} also indicate that, there exists $\bar{d}>0$ and $\tilde{T}$ such that $\bar{d} \ge \frac{2NL}{T_1+2} > |u(f_{T_1+1})- u(f_{T_1})|$. Consequently, there exists $\bar{d}>0$, that implies $u(\underline{\sigma}^{*(m')})> u(f_{T'_1+1}) > u(\bar{\sigma}^{*(m'')})$. However, by Lemma \ref{lem_inc}, if $f_t$ is outside the set $\Sigma_{N\delta+\epsilon_1}$, the potential value has to increase until again entering into the approximate equilibrium set. This creates a contradiction, since the maximum potential $u(\bar{\sigma}^{*(m'')})$ around another equilibrium $a^{*(m'')}$ is less than $u(f_t)$. Thus, this holds for any equilibrium pair, and guarantees that the sequence $f_t$ only visits around a single equilibrium after long time enough $t>T$. 
\end{proof}


Theorem \ref{lem_step4} is a sequel to the results in \cite{aydin2021decentralized}. DFP creates additional error rate of  $O(\log t/t^2)$ in \eqref{eq_step_inc}, while FP has an error rate of $O(1/t^2)$ \cite{candogan2013dynamics}. Since in both cases, the error rate goes to $0$, DFP  recovers the convergence results of FP with the same set of assumptions on the game structure.
\begin{figure}
	\centering
	\begin{tabular}{cc}
	\includegraphics[width=.5\linewidth]{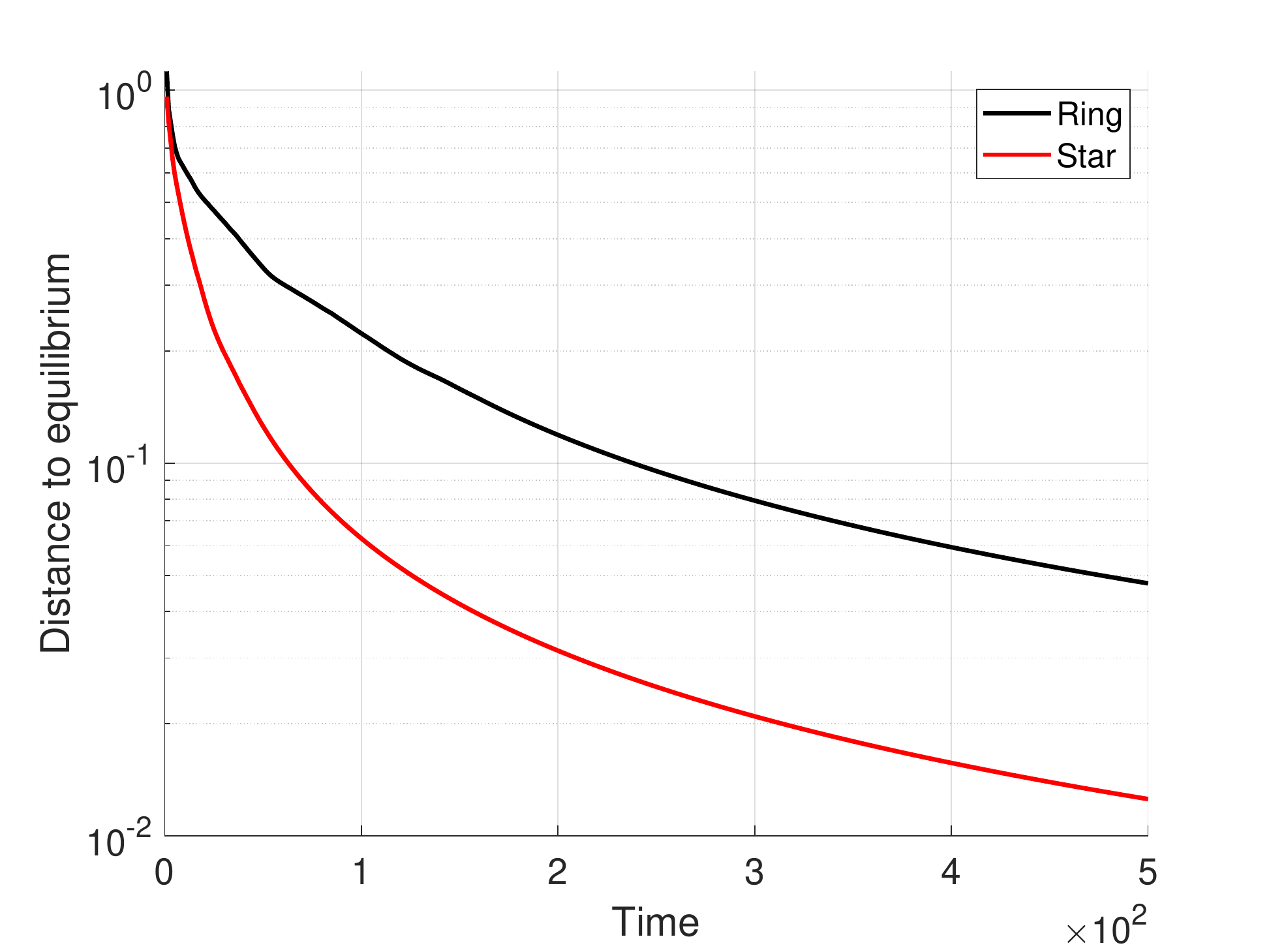}
	\includegraphics[width=.5\linewidth]{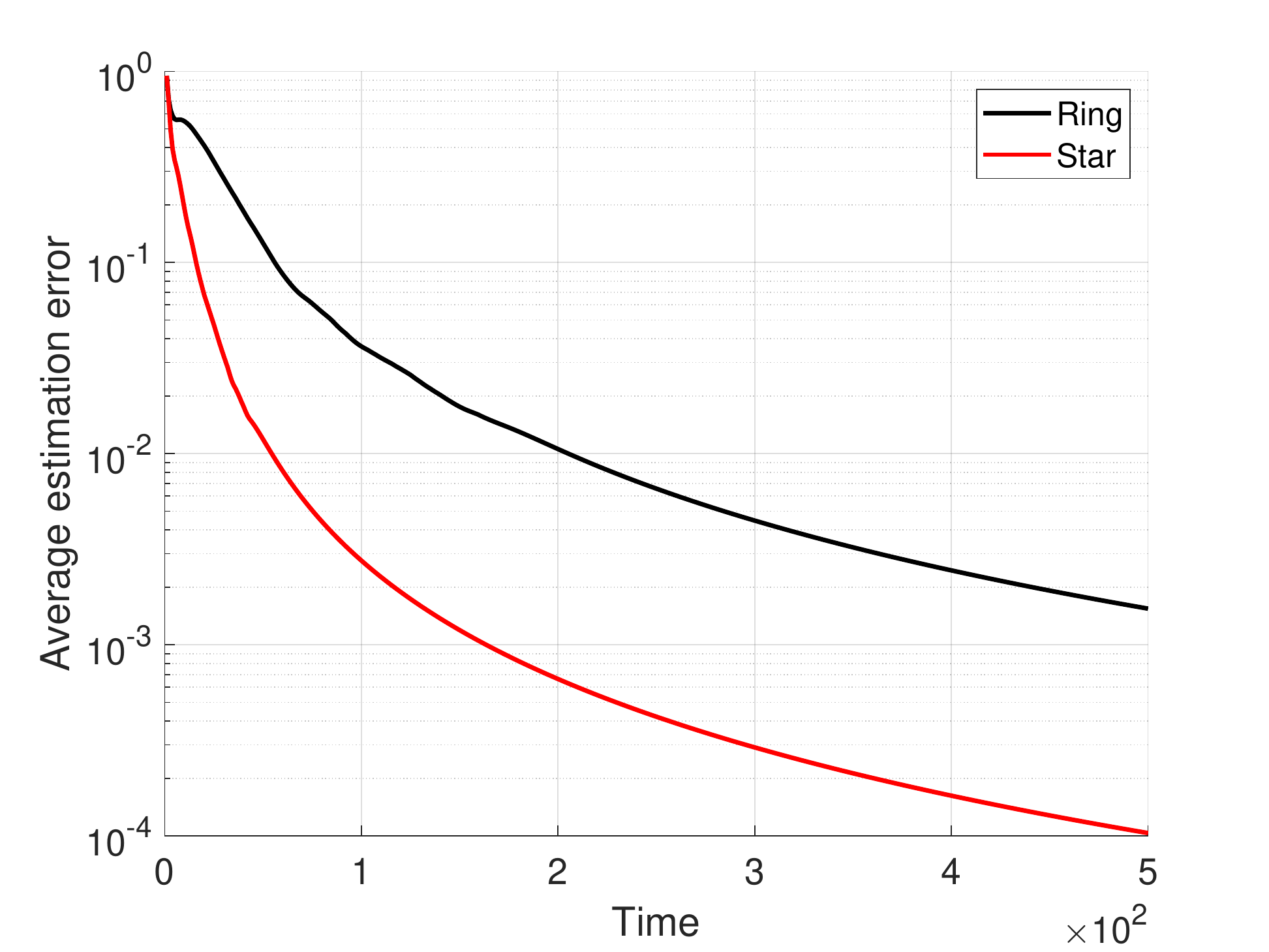}
	\end{tabular}
	\caption{DFP in target assignment game with unknown target locations over 20 runs. (Left)   Average distance to Nash equilibrium $\frac{1}{N}\sum_{i \in \ccalN} || f_{it}-\sigma^*_i||$ (Right) Average estimation error $\frac{1}{N(N-1)}\sum_{i \in \ccalN} \sum_{j \in \ccalN \setminus \{i\}}||f_{it}-\upsilon_{it}^j ||$. } \vspace{-1pt}
	\label{fig_conv}
	\vspace{-12pt}
\end{figure}
\section{Numerical Experiments}\label{sec::numeric}


We use target assignment game with $N=10$ autonomous agents and $K=10$ targets in the numerical experiments. Each agent $i \in \ccalN$ selects only one target $k \in  \ccalK:=\{1,\cdots,K\}$ so that the objective is to cover all targets with minimum effort as a team, given the utility functions defined as below,
\begin{equation}\label{util_target}
    u_{i}(a_i,a_{-i})= \frac{a_i^T\mathbb{1}_{a_{-ik}=0}}{a_i^T d_{i}},
\end{equation}
where $a_i=\bbe_k \in \mathbb{R}^K$ is an unit vector and $\mathbb{1}_{a_{-ik}=0} \in \{0,1\}^K$ is a binary vector whose $k^{th}$ index is 1 if none of the other agents $j \in \ccalN \setminus \{i\}$ select $k$, and otherwise the $k^{th}$ index is equal to 0. The distance vector  $d_i=[d_{i1},\cdots,d_{ik}, \cdots, d_{iK}] \in \mathbb{R}_+^K $ captures the distance between targets and  agent $i$. We assume that target locations are unknown, and agents obtain noisy private signals about target locations coming from a normal distribution with the mean equal to the target locations and variance $0.1$ for each dimension ($x$ and $y$ axes) of the target location independently. Then, agents take sample average of private signal for targets coming at each time step $t$. Agents stop receiving these signals after time step $t=10$. Target assignment game with equal distances between targets and agents is a potential game. We convert it into a near-potential game by introducing different estimates and distances to targets. 
We use ring and stars communication networks. Self-weights are set as $w_{i,l}^i=0.75$ while $w^i_{j,l}=0.25/|\ccalN_i|, \forall{j} \in \ccalN_i$ is selected for neighbor agents.

We implemented $20$ runs of numerical experiments with randomly created distances.  Fig.~\ref{fig_conv} (Right) shows the average estimation error between local and real empirical frequencies. It corroborates the convergence rate of $O(\log t/t)$ given in Lemma \ref{lem_inc}. Further, we observe that agents' final action profiles converge to an one-to-one assignment between agents and targets in all cases. The star network converges faster than the ring network.  Fig.~\ref{fig_conv} (Left) shows the average rate of convergence of the empirical frequencies to a NE of the target assignment game which is an one-to-one assignment of agents to targets. It also confirms the result of convergence around a single equilibrium point. 

\section{Conclusion}
In this paper, we studied the convergence of DFP in near-potential games. We proved empirical frequencies of actions converge to an approximate NE around a single NE point. That is, DFP preserves the convergence properties of FP, despite the lack of perfect information on others' past actions. This result implies that decentralized and repetitive best-response type behavior in large-scale networked systems, converge to rational behavior.
\bibliographystyle{IEEEtran}
\bibliography{bibliography}

\end{document}

%% file: my_sections.tex
\usepackage{needspace}



%% file: introduction.tex
Game theory deals with systems having multiple decision-makers. In non-cooperative games, agents take actions to maximize their individual utility functions that depend on the actions of other agents. Potential games is a special class of games that capture scenarios where there exists a common function modeling the change in individual utilities, named as potential function. Applications of potential games appear in various large-scale networked systems including transportation systems \cite{marden2009joint}, mobile robotic systems \cite{eksin2017distributed}, and communication networks \cite{candogan2010near}. Decentralized decision-making protocols, e.g., best-response \cite{monderer1996potential, swenson2018distributed}, fictitious play (FP) \cite{monderer1996fictitious,swenson2017single}, are used to understanding emerging behavior or to design individual actions in such large-scale systems. 
A common assumption in the convergence of these protocols is that agents have full or common information about their utility functions or the potential function. Here, we lift this assumption by allowing the game agents are playing to deviate from an exact potential game.

Near-potential games \cite{candogan2013dynamics} extend  potential games, by defining games as a deviation from a potential game. This deviation may stem from incomplete information about payoff-relevant environment parameters. 
Specifically, the deviation between two games is defined in terms of unilateral change of actions, where only one agent changes its and others stay in the same profile. If this deviation is bounded, traditional decision-making protocols, e.g., best-response, or FP, converge to a region around NE, i.e., an approximate-NE \cite{candogan2013dynamics}. In this paper, we analyze convergence properties of a decentralized version of FP (DFP) where agents can only exchange information with a subset of their neighbors after each decision epoch, which is in contrast to standard FP that assumes agents have perfect information about the past actions of other agents.

In particular, we consider agents taking actions with respect to DFP in time-varying communication networks as in \cite{aydin2021decentralized}. In \cite{aydin2021decentralized}, we had shown that DFP converges to a set of strategies that obtain potential function values comparable to the set of approximate Nash equilibria. Here, we extend these results to show that empirical frequencies of agents converge around a single NE (Theorem \ref{lem_step4}) given two additional assumptions: {\it i)} number of Nash equilibria is finite, and {\it ii)} the near-potential game is close enough to a potential game. Numerical experiments on a target assignment game with unknown payoffs show that the action profiles can actually converge to the exact NE of the closest potential game (the game with known payoffs). Together these results show that DFP can be used to model or design team behavior in large-scale networked systems where agents communicate over a time-varying network, and have different information about a given common goal.